\documentclass[aps,prd,superscriptaddress,nofootinbib,preprint]{revtex4}
	\usepackage{amsmath,amssymb}
	\usepackage{makeidx}
	\usepackage{amsfonts}
	\usepackage[ansinew]{inputenc}
	\usepackage[usenames,dvipsnames]{pstricks}
	\usepackage{subfigure}
	\usepackage{epsfig}
	\usepackage{pst-grad} 
	\usepackage[colorlinks,hyperindex]{hyperref}
	\hypersetup
	{
		colorlinks,%
		citecolor=black,%
		linkcolor=black,%
		urlcolor=black,%
	}
    \usepackage{setspace}
	
\newcounter{concno}

\newcommand{\be}{\begin{equation}}
\newcommand{\ee}{\end{equation}}
\newcommand{\bea}{\begin{eqnarray}}
\newcommand{\eea}{\end{eqnarray}}

\newcommand{\ba}{\begin{array}}
\newcommand{\ea}{\end{array}}

\newcommand{\refer}[1]{(\ref{#1})}
\newcommand{\dd}{\mbox{d}}

\begin{document}

\title{On the Possibility to Determine Neutrino Mass Hierarchy via Supernova Neutrinos with Short-Time Characteristics}
\date{\today}

\author{Junji Jia}

\email{junjijia@whu.edu.cn}

\affiliation{School of Physics and Technology, Wuhan University, Wuhan 430072, China}

\affiliation{Center for Theoretical Physics, Wuhan University, Wuhan 430072, China}

\author{Yaoguang Wang}

\email{yaoguang.wang@whu.edu.cn}

\affiliation{School of Physics and Technology, Wuhan University, Wuhan 430072, China}

\affiliation{Institute of High Energy Physics, Chinese Academy of Sciences, Beijing 100049, China}

\author{Shun Zhou}

\email{zhoush@ihep.ac.cn}

\affiliation{Institute of High Energy Physics, Chinese Academy of Sciences, Beijing 100049, China}

\affiliation{School of Physical Sciences, University of Chinese Academy of Sciences, Beijing 100049, China}

\affiliation{Center for High Energy Physics, Peking University, Beijing 100871, China}

\begin{abstract}
In this paper, we investigate whether it is possible to determine the neutrino mass hierarchy via a high-statistics and real-time observation of supernova neutrinos with short-time characteristics. The essential idea is to utilize distinct times-of-flight for different neutrino mass eigenstates from a core-collapse supernova to the Earth, which may significantly change the time distribution of neutrino events in the future huge water-Cherenkov and liquid-scintillator detectors. For illustration, we consider two different scenarios. The first case is the neutronization burst of $\nu^{}_e$ emitted in the first tens of milliseconds of a core-collapse supernova, while the second case is the black hole formation during the accretion phase for which neutrino signals are expected to be abruptly terminated. In the latter scenario, it turns out only when the supernova is at a distance of a few Mpc and the fiducial mass of the detector is at the level of gigaton, might we be able to discriminate between normal and inverted neutrino mass hierarchies. In the former scenario, the probability for such a discrimination is even less due to a poor statistics.
\end{abstract}


\maketitle

\section{Introduction}

The fact that neutrinos have finite and non-degenerate masses has well been established by a number of elegant neutrino oscillation experiments~\cite{Fukuda:1998mi, Fukuda:2001nj, Ahmad:2001an, Ahmad:2002jz, Eguchi:2002dm, Michael:2006rx, Abe:2011sj, Adamson:2011qu, An:2012eh, Ahn:2012nd, Abe:2012tg}. Thanks to these successful experiments, we currently know three neutrino flavor mixing angles $\theta^{}_{12} \approx 34^\circ$, $\theta^{}_{23} \approx 45^\circ$ and $\theta^{}_{13} \approx 9^\circ$, and two neutrino mass-squared differences $\Delta m^2_{21} \equiv m^2_2 - m^2_1 \approx 7.53\times 10^{-5}~{\rm eV}^2$ and $\Delta m^2_{32} \equiv m^2_3 - m^2_2 \approx 2.45\times 10^{-3}~{\rm eV}^2$ for the normal neutrino mass hierarchy (NH), or $\Delta m^2_{23} \equiv m^2_2 - m^2_3 \approx 2.52\times 10^{-3}~{\rm eV}^2$ for the inverted neutrino mass hierarchy (IH)~\cite{Olive:2016xmw}. Whether NH or IH is realized in nature is a fundamentally important question that may affect not only theoretical model building for neutrino mass generation and lepton flavor mixing, but also a variety of future neutrino experiments aiming to discover the leptonic CP violation, to probe the absolute scale of neutrino masses and to pin down their Dirac or Majorana nature.

To determine the neutrino mass hierarchy (MH), many methods by implementing accelerator neutrinos~\cite{Ishitsuka:2005qi}, reactor neutrinos~\cite{Petcov:2001sy, Choubey:2003qx, Learned:2006wy, Zhan:2008id, Zhan:2009rs, Li:2013zyd}, atmospheric neutrinos~\cite{Nunokawa:2005nx, Ribordy:2013set} or supernova neutrinos~\cite{Dighe:1999bi, Lunardini:2003eh, Dighe:2003be, Duan:2007bt, Dasgupta:2008my, Serpico:2011ir, Lai:2016yvu, Scholberg:2017czd,Kyutoku:2017wnb} have been proposed. Among them, those methods using supernova (SN) neutrinos are particularly interesting because of the interplay between intrinsic properties of massive neutrinos and the long-sought mechanism of SN explosions. Historically, the detection of neutrinos from SN1987A in the Large Magellanic Cloud~\cite{Hirata:1987hu, Bionta:1987qt} has motivated a huge amount of theoretical works in both SN physics and neutrino physics. Due to these potentials and gains from past experience, all modern running and forthcoming neutrino observatories, such as IceCube~\cite{Aartsen:2014njl}, Hyper-Kamiokande~\cite{Abe:2011ts}, JUNO~\cite{An:2015jdp} and KM3NeT~\cite{Adrian-Martinez:2016fdl} have set SN neutrino detection as one of their main physics targets.

Now that neutrinos are massive particles, it is quite evident that for two neutrino mass eigenstates with masses $m^{}_i$ and $m^{}_j$ respectively, the time difference $\Delta t^{}_{ij} \equiv t^{}_i - t^{}_j$ for travelling from the SN to the detector is~\cite{Zatsepin:1968kt}
\be
\Delta t_{ij}=5.15~\mbox{ms}\cdot \frac{\Delta m_{ij}^2/\mbox{eV}^2}{\left(\langle E\rangle/\mbox{10~MeV}\right)^2} \cdot \frac{D}{\mbox{10~kpc}} \; , \label{tof}
\ee
where $\Delta m^2_{ij} \equiv m^2_i - m^2_j$ denotes the neutrino mass-squared difference, $\langle E\rangle$ is the average neutrino energy and $D$ is the distance of the SN. Note that the energies of two neutrino mass eigenstates in Eq.~(\ref{tof}) have been assumed to be the same and represented by the average neutrino energy $\langle E \rangle \gg m^{}_i$ to estimate the difference in the time of flight (ToF). In this work, we show that SN neutrino emission with a characteristic time smaller than the difference in the ToF for different neutrino mass eigenstates can be utilized to determine the neutrino MH, if the time resolution of the future SN neutrino detector is good enough as expected. For illustration, we consider two distinct scenarios in which these features of SN neutrino signals could be satisfied.

The first scenario is the short burst of electron neutrinos in the early-time process of neutronization (i.e., $e^- + p \to n + \nu^{}_e$) during supernova explosion, which takes place right after the prompt shock wave forms at the surface of inner core and starts to disassociate heavy nuclei in the surroundings. The existence of such a neutronization burst has been found to be a robust feature of SN neutrino emission in all the numerical simulations of SN explosions using different equations of state, treatment of gravity, and numerical approaches for hydrodynamics~\cite{Ott:2012mr, Couch:2013kma, Takiwaki:2013cqa, Bruenn:2012mj, Mezzacappa:2000jb}. Depending on the details of simulations, the time duration of $\nu^{}_e$ burst, characterized by its full width at half maximum of the luminosity and denoted by $\Delta t^{}_{\rm N}$, can range from $3$ ms to $20$ ms. A typical value of $\Delta t^{}_{\rm N} = 5$ ms, which is close to the median of all the possible values that we have surveyed \cite{Liebendoerfer:2002xn, Marek:2005if, Lentz:2011aa, Hempel:2011mk,Kuroda:2012nc, Buras:2005rp, Buras:2005tb, Liebendoerfer:2003es, Fischer:2008rh, Serpico:2011ir}, will be adopted in the following discussions.

The second scenario is the abrupt termination of neutrino emission due to the formation of a black hole (BH) in the accretion phase of a core-collapse SN. The BH formation is expected to occur when the mass of the progenitor star is between about 25 and 40 solar masses~\cite{Woosley:2002zz} and the shock wave cannot manage to successfully propagate out of the heavy outer core even with neutrino heating. Although there will be no final SN explosion, this scenario does have the advantage of an even shorter characteristic time, i.e., $\Delta t^{}_{\rm B}\approx 2R/c\sim 0.1$ ms, where the radius of active region for neutrino emission $R \approx 10~{\rm km}$ and the speed of light $c \approx 3\times 10^{10}~{\rm cm}~{\rm s}^{-1}$ have been used. At this point, we should mention that the ToF of massive neutrinos from the BH forming SNe have been considered previously by Beacom {\it et al.} in Ref.~\cite{Beacom:2000qy}, where an upper bound on absolute neutrino masses has been obtained.

Different from previous works, the present paper will concentrate on the impact of ToF on the time distribution of SN neutrino events in future detectors. In particular, we explore the distinct features in the cases of NH and IH. In Ref.~\cite{Serpico:2011ir}, the rising time of SN neutrino event rate in IceCube has been implemented to discriminate between NH and IH. Other possibilities have recently been summarized in Ref.~\cite{Scholberg:2017czd}. The analysis presented in our paper can be regarded an additional effort in the same direction. The remaining part is organized as follows. In Sec.~\ref{secburst}, the time distribution of SN $\nu^{}_e$-burst events in the water-Cherenkov and liquid-scintillator detectors is predicted by taking account of the difference in ToF of different neutrino mass eigenstates, while Sec.~\ref{secbh} is devoted to the tail of neutrino events in the case of BH formation. Finally, we conclude in Sec.~\ref{secdis}.

\section{The neutronization burst\label{secburst}}

For the neutronization burst, the neutrino flavor state $|\nu^{}_e\rangle$ is produced and then propagating from the SN to the detector on the Earth. During the propagation of such a long distance (e.g., $D = 51~{\rm kpc}$ for SN1987A), the coherence will be lost and the narrow peak of $|\nu^{}_e\rangle$ will split into those of three neutrino mass eigenstates $|\nu^{}_i\rangle$ (for $i = 1, 2, 3$) due to the difference in arrival time according to Eq.~\refer{tof}. Because the separation of these peaks depends on the MH in a simple and definitive way, we can immediately recognize the MH once SN neutrinos are detected with high enough statistics.

\subsection{General Remarks \label{nbgen}}

First of all, we give some general remarks on the time separation of three possible peaks and its dependence on the SN distance $D$, neutrino MH and neutrino flavor conversions. Since the time resolution of most modern neutrino detectors are at the $\mathcal{O}$(10)~ns level, which is accurate enough to reconstruct the peak with a width about $\Delta t^{}_{\rm N} \approx 5~{\rm ms}$, the only requirement to temporally resolve the peaks of different neutrino mass eigenstates is that the difference in arrival time is larger than the peak width, i.e., $|\Delta t^{}_{ij}| \geq \Delta t^{}_{\rm N}$. Given the neutrino energy, this inequality can be translated into a lower bound on the SN distance
\be
D \geq \frac{\Delta t^{}_{\rm N}}{5.15~\mbox{ms}} \cdot \frac{\left(\langle E\rangle /\mbox{10~MeV}\right)^2}{|\Delta m_{ij}^2|/\mbox{eV}^2}\cdot 10~\mbox{kpc}\equiv D_{\mbox{\scriptsize Nij}} \; , \label{dmndef}
\ee
where we define $D_{\mbox{\scriptsize Nij}}$ to be the minimal resolvable distance. Then, corresponding to $\Delta m^2_{21}$ and $|\Delta m^2_{32}|$, there will be two independent distances $D^{}_{\mbox{\scriptsize N21}}$ and $D^{}_{\mbox{\scriptsize N32}}$, satisfying the condition $D^{}_{\mbox{\scriptsize N21}} \gg D^{}_{\mbox{\scriptsize N32}}$ due to $\Delta m^2_{21} \ll |\Delta m_{32}^2|$ from neutrino oscillation data.

Some discussions on the SN distance are in order. In the case of large distances $D > D_{\mbox{\scriptsize N21}}$, for both neutrino MHs, there will be three well separated peaks with different ToFs and event rates, each corresponding to one of neutrino mass eigenstates. For the NH, there should first come into the detector two peaks of $|\nu^{}_1\rangle$ and $|\nu^{}_2\rangle$ at time $t^{}_1$ and $t^{}_2$, respectively, with a difference of $\Delta t^{}_{21} = t^{}_2 - t^{}_1 > 0$. Then after a relatively longer time $\Delta t_{32}$, which according to Eq.~\refer{tof} should be $\Delta m^2_{32}/\Delta m^2_{21}$ times longer than $\Delta t_{21}$,  there appears the third peak, corresponding to $|\nu^{}_3\rangle$. For the IH, the peak corresponding to $|\nu^{}_3\rangle$ will reach the detector earlier than the $|\nu^{}_1\rangle$ and $|\nu^{}_2\rangle$ peaks do, with the order of the latter two being the same as that in the NH. Therefore, even though the size of each of the time differences $\Delta t_{21}$ and $|\Delta t_{32}|$ are the same for two neutrino MHs, their temporal order appearing in the detector leads to a clear and unique signature of the MH. Once this ordering is experimentally observed, the MH will be determined completely. In contrast, if the SN distances turn out to be small $D < D^{}_{\mbox{\scriptsize N32}}$, then all three peaks become indistinguishable, and it will be impossible to deduce the neutrino MH from the appearing order of the peaks.

If the distance is lying in between $D_{\mbox{\scriptsize N32}}$ and $D_{\mbox{\scriptsize N21}}$, the peak corresponding to the mass eigenstate $|\nu^{}_3\rangle$ will be discriminable from the peaks of $|\nu^{}_2\rangle$ and $|\nu^{}_1\rangle$, while the latter two are non-separable and we will effectively see only two peaks. Therefore, the method utilizing the temporal order of the observed peaks does not work in this case. However, we can make use of another piece of information, namely, the magnitude of event rate $R^{}_i$ at those peaks corresponding to each neutrino mass eigenstate $|\nu^{}_i\rangle$ (for $i = 1, 2, 3$).
To study the event rates of these peaks at the detector, one needs to know not only the initial $|\nu_e^{}\rangle$ spectrum and the reaction for detection, but also how the $|\nu^{}_e\rangle$ spectrum evolves when neutrinos propagating from the production region to the surface of the SN.

\subsection{Flavor Conversions \label{nbosc}}

The short-time burst of $|\nu_e^{}\rangle$ is generated when the first strike of the outer materials onto the inner SN core is bounced back and the heavy nuclei are disintegrated into free nucleons by the energetic shock wave~\cite{Mirizzi:2015eza, Janka:2017vlw}. During the outward propagation of SN neutrinos from the dense region to the surface, it is generally expected that both the ordinary Mikheyev-Smirnov-Wolfenstein (MSW) matter effects~\cite{Mikheev:1986gs, Wolfenstein:1977ue} and neutrino self-induced collective oscillations~\cite{Pantaleone:1992xh, Samuel:1993uw, Duan:2005cp, Duan:2006an} will be crucially important for neutrino flavor conversions. See, e.g., Refs.~\cite{Duan:2010bg, Mirizzi:2015eza, Chakraborty:2016yeg}, for recent reviews on collective oscillations of SN neutrinos.

For the progenitor stars of more than 10 solar masses, they will normally develop heavy iron cores before collapse. In this case, it is natural to expect that during explosion, collective neutrino oscillations occur within a few hundred kilometers above SN core, whereas the MSW effects will come into play far away in the SN envelope. However, for the stars of 8 to 10 solar masses, which finally evolve to the O-Ne-Mg core-collapse supernovae, the matter density profile above the core is so steep that the MSW resonances could happen within the region of collective oscillations~\cite{Duan:2007sh, Dasgupta:2008cd, Cherry:2010yc}. For supernovae that allow only MSW effect for the neutronization burst, from Ref.~\cite{Dighe:1999bi}, we see that in the NH case the initially generated neutronization $|\nu_e^{}\rangle$ burst  is equivalent to the heaviest mass eigenstate  $|\nu^{\rm m}_3\rangle$ in matter. Due to a relatively large $\theta_{13}$~\cite{An:2012eh}, the flavor conversion proceeds adiabatically when $|\nu^{}_e\rangle$ passes through the density regions of $\Delta m^2_{32}$- and $\Delta m^2_{21}$-driven resonances and finally becomes a mass eigenstate $|\nu^{}_3\rangle$ in vacuum. For the IH case, the initial $|\nu_e^{}\rangle$ burst is almost a mass eigenstate $|\nu^{\rm m}_2\rangle$ in matter. Again this neutrino state will also traverse the entire density profile adiabatically and become the mass eigenstate $|\nu^{}_2\rangle$ in vacuum after emerging.
On the other hand, for supernovae that allow not only the MSW but also the collective
oscillations, the initial neutronization $|\nu_e\rangle$ burst usually evolves to all three mass eigenstates after emerging
with different probabilities. Ref. \cite{Dasgupta:2008cd} have given transition probabilities of $|\nu_e\rangle \to |\nu_i\rangle$ as a function of energy
\be \mbox{Prob}(|\nu_e\rangle\to |\nu_i\rangle,E)\equiv P_{ei}(E) \label{eqprob1}\ee
in its Fig. 2. We will use this $P_{ei}(E)$ for the neutronization $|\nu_e\rangle$ burst in supernova allowing the collective oscillation.

Although the self-induced collective oscillation has been proposed, it remains unclear to what extent will this occur in a real SN environment due to the large uncertainties such as progenitor mass, neutrino luminosity and simulation details including dimensionality and multi-angular/single-angular technique in dealing with collective oscillation.
Also noticing that the researches on this topic are rapidly advancing, therefore we will remain conservative in this work by considering two different cases for the conversion probability from initial $|\nu_e\rangle $ to $|\nu_i\rangle$ after emerging:
\begin{itemize}
\item {\it Case (A)} -- For the O-Ne-Mg core-collapse SNe, both MSW effects and neutrino self-induced collective oscillations play an important role, so the initial neutronization $|\nu_e^{}\rangle$ burst usually evolves to all three mass eigenstates in vacuum but with different probabilities. As already demonstrated in Ref.~\cite{Dasgupta:2008cd}, the transition probabilities for $|\nu^{}_e\rangle \to |\nu^{}_i\rangle$ (for $i = 1, 2, 3$) are actually functions of the neutrino energy, i.e., $P^{}_{ei}(E)$, where the spectral splits are found and explained analytically. In the following calculations, we use the probabilities from Ref.~\cite{Dasgupta:2008cd} as the first example.

\item {\it Case (B)} -- Second, as a simple working assumption, we neglect the energy dependence and specify the transition probabilities for $|\nu^{}_e\rangle \to |\nu^{}_i\rangle$ as $P^{}_{e1} = 1/6$, $P^{}_{e2} = 1/3$ and $P^{}_{e3} = 1/2$, representing a class of scenarios in which the probabilities are comparable in magnitude. Although the exact values of those probabilities are not important, the discriminating power for MH will be lost if any one of them becomes negligibly small.
\end{itemize}

It is worthwhile to mention that there are large uncertainties in the progenitor mass, neutrino luminosity and the details of numerical simulations, such as dimensionality, equations of state and neutrino transport, so it is a complicated situation to deal with collective oscillations. In the following, we will compute the event rates of three neutrino mass eigenstates only in the above two cases for illustration.

\subsection{Neutrino Event Rates \label{nber}}

To numerically check whether the neutronization burst method is feasible in currently running and future detectors, we calculate neutrino event rates. Starting with the neutrino spectrum $n^{}_{\nu_e^{}}$, we will adopt the quasi-thermal spectrum approximated by a Gamma distribution~\cite{Keil:2002in, Tamborra:2012ac, Tamborra:2014hga}
\bea
n^{}_{\nu_e^{}}(t, E) = \frac{L(t)}{\langle E(t)\rangle} \frac{E^\alpha}{\Gamma(\alpha+1)}\left(\frac{\alpha+1}{\langle E(t)\rangle}\right)^{\alpha+1} \exp\left(-\frac{(\alpha+1)E}{\langle E(t)\rangle}\right) \; , \label{nbn}
\eea
where the spectral index $\alpha$ is given by
\be
\alpha = \frac{\langle E(t)\rangle - E_{\mbox{\scriptsize rms}}(t)^2}{E_{\mbox{\scriptsize rms}}(t)^2} > 2 \; ,
\ee
$\langle E(t)\rangle$ is the average energy, and $E_{\mbox{\scriptsize rms}}(t)$ is the root-mean-square, while the luminosity $L(t)$ is given by the SN simulation data~\cite{Tamborra:2014hga}. Since the burst of $|\nu^{}_e\rangle$ will finally evolve into an incoherent superposition of three neutrino mass eigenstates $|\nu^{}_i\rangle$ at the SN surface, the number densities of the latter are given by
\be
n^{}_{\nu_i}(t, E) = n^{}_{\nu_e^{}}(t, E) P^{}_{ei}(E) \; ,
\ee
where the transition probabilities $P^{}_{ei}(E)$ have been specified in Sec.~\ref{nbosc} and the separation of neutrino mass eigenstates within the SN has safely been neglected. After emerging from the SN surface, three mass eigenstates propagate towards Earth and arrive in the detector at different time due to the long distance. At the detector, if all three mass eigenstates are received at some time instant $t^{}_{\rm d}$, then $|\nu^{}_2\rangle$ should have left the supernova earlier by $\Delta t^{}_{21}$ than $|\nu^{}_1\rangle$, while $|\nu^{}_3\rangle$ should have left earlier (or later) by $\Delta t_{32}$ (or $\Delta t^{}_{23}$) than $|\nu_2\rangle$ in the NH (or IH) case. Hence the flux $F_{\nu_i}(t_d,E)$ for $|\nu_i\rangle$ at the detector and the time $t^{}_{\rm d}$ becomes
\bea
F^{}_{\nu_i}(t^{}_{\rm d}, E) = \frac{1}{4\pi D^2} n_{\nu_e^{}}(t^{}_{\rm d} - \Delta t^{}_{i1}, E) P^{}_{ei}(E) \; ,\label{fcalc}
\eea
where we have dropped a common ToF of $|\nu^{}_1\rangle$ and identified the detection time $t^{}_{\rm d}$ as the the emission time of  $|\nu^{}_1\rangle$ without loss of any generality.

For SN neutrinos of energies about a few tens of MeV, both elastic neutrino-proton and neutrino-electron scatterings can be observed in the liquid-scintillator detectors due to a low energy threshold, while only the elastic neutrino-electron scattering is observable in the water-Cherenkov detectors.  Moreover, for liquid-scintillator detector, it allows charge-current interactions and neutral-current interactions with its $^{12}$C nuclei. The differential cross section for elastic neutrino-proton scattering is universal for all neutrino flavors at the lowest order and given by~\cite{Weinberg:1972tu, Beacom:2002hs}
\be
\frac{\dd \sigma_{\nu p}}{\dd T^{}_p} = \frac{G_{\rm F}^2 M^{}_p}{\pi}\left[\left(1-\frac{M^{}_p T^{}_p}{2 E_\nu^2}\right)c_{\rm V}^2+ \left(1+\frac{M^{}_p T^{}_p}{2E_\nu^2}\right) c_{\rm A}^2\right] \; , \label{nuallpcs}
\ee
where $G^{}_{\rm F} = 1.166\times 10^{-5}~{\rm GeV}^{-2}$ is the Fermi constant, $M^{}_p$ and $T^{}_p$ are the proton mass and kinetic energy, respectively, and the vectorial and axial couplings are $c^{}_{\rm V} = (1 - 4\sin^2\theta_{\rm W})/2$ and $c^{}_{\rm A} = 1.27/2$ with the Weinberg angle $\sin^2 \theta^{}_{\rm W} \approx 0.23$. Assuming the lepton flavor mixing matrix to be unitary, one can obtain the event spectrum corresponding to the mass eigenstate $|\nu^{}_i\rangle$ as follows
\be
r^{}_{i,{\rm PS}}(t^{}_{\rm d}, E^{}_\nu) = N^{}_p \int_{T^{}_{\rm c}}^{2E_\nu^2/M^{}_p} F^{}_{\nu^{}_i}(t^{}_{\rm d}, E^{}_\nu) \cdot \frac{\dd \sigma^{}_{\nu p}}{\dd T^{}_p} \cdot \dd T_p \label{eratediff}\ee
where $N^{}_p$ is the total number of protons in the target. For the liquid-scintillator detectors, the recoil energy of the final-state proton will be quenched significantly and one can establish the relationship between the original recoil energy and the observed energy as done in Refs.~\cite{Beacom:2002hs, Dasgupta:2011wg, Lu:2016ipr}. Furthermore, when the observed energy falls below 0.2 MeV, the radioactive backgrounds will be dominant. Therefore, we have placed an energy cut at $T^{}_{\rm c} = 0.2~{\rm MeV}$, corresponding to an original recoil energy of $T^{}_p = 1.0~{\rm MeV}$, for which a minimal neutrino energy $E^{\rm min}_\nu = (T^{}_p M^{}_p/2)^{1/2} \approx 21.7~{\rm MeV}$ is required.

As we have mentioned, the recoil energy of proton under discussions is at most a few MeV and thus there will be no Cherenkov light, so it is impossible to observe any signals from neutrino-proton scattering in the water-Cherenkov detectors. For the elastic neutrino-electron scattering, the total cross sections for $\nu^{}_e$ and $\nu^{}_x$, where the latter collectively denotes $\nu^{}_\mu$ and $\nu^{}_\tau$ and their antiparticles, are well known in the standard model. At the tree level, the explicit expressions are~\cite{tHooft:1971ucy, Marciano:2003eq}
\bea
\frac{\dd\sigma^{}_{\nu e}(E^{}_\nu)}{{\rm d}T^{}_e} &=& \frac{2 m^{}_e G^2_{\rm F}}{\pi} \left[\epsilon^2_- + \epsilon^2_+ \left(1 - \frac{T^{}_e}{E^{}_\nu}\right)^2 - \epsilon^{}_- \epsilon^{}_+ \frac{m^{}_e T^{}_e}{E^2_\nu}\right] \; ,  \label{nuxecs}
\eea
where the kinetic energy of the final-state electron $T^{}_e \equiv E^\prime_e - m^{}_e$ is lying below $T^{\rm max}_e = E^{}_\nu/[1 + m^{}_e/(2E^{}_\nu)]$. For electron neutrinos, the coefficients are given by $\epsilon^{}_- = -1/2 - \sin^2 \theta^{}_{\rm W}$ and $\epsilon^{}_+ = - \sin^2 \theta^{}_{\rm W}$; while $\epsilon^{}_- = 1/2 - \sin^2 \theta^{}_{\rm W}$ and $\epsilon^{}_+ = -\sin^2 \theta^{}_{\rm W}$ for muon and tau neutrinos.
Thus, the event spectrum of neutrino mass eigenstate $|\nu_i\rangle$ can be simply written as
\be
r^{}_{i,{\rm ES}}(t^{}_{\rm d}, E^{}_\nu) = N^{}_e \sum_\beta \int^{T^{\rm max}_e}_{T^{}_{\rm c}} F^{}_{\nu_i}(t^{}_{\rm d}, E^{}_\nu)\cdot |U^{}_{\beta i}|^2 \cdot \frac{\dd \sigma^{}_{\nu^{}_\beta e}}{\dd T^{}_e} \cdot \dd T^{}_e \; ,
\label{neratediff}
\ee
where $N^{}_e$ is the total number of electrons in the target and $|U^{}_{\beta i}|^2$ is the probability for the projection of $|\nu^{}_i\rangle$ to $|\nu^{}_\beta\rangle$ with $U^{}_{\beta i}$ being the lepton flavor mixing matrix. For the neutrino reaction with $^{12}$C, the charge-current interactions
\be
\nu_e~+~^{12}\mbox{C}\to e^-~+~^{12}\mbox{B}, ~~\bar{\nu}_e~+~^{12}\mbox{C}\to e^+~+~^{12}\mbox{N}, \label{c12cc}\ee
and neutral current interactions
\be
\nu_i~+~^{12}\mbox{C}\to \nu_i^\prime~+~^{12}\mbox{C}^*, ~~\bar{\nu}_i~+~^{12}\mbox{C}\to \bar{\nu}_i^\prime~+~^{12}\mbox{C}^* \label{c12nc}\ee
have been well established both theoretically and experimentally. In this work, for liquid-scintillator neutrino detectors, we will consider these reaction too. We will directly use the cross-section tabulated in Tab. 1 of Ref. \cite{Fukugita:1988hg}. Denoting these cross-section as
\be
\sigma_{\nu_\beta\to ^{12}\mathrm{C}^*}(E_{\nu_\beta}),~ \sigma_{\bar{\nu}_\beta\to ^{12}\mathrm{C}^*}(E_{\bar{\nu}_\beta}),~ \sigma_{\nu_e\to ^{12}\mathrm{N}}(E_{\nu_e}),~ \sigma_{\bar{\nu}_e\to ^{12}\mathrm{B}}(E_{\bar{\nu}_e}),\ee
respectively, we can similarly compute the event spectrum of neutrino mass eigenstate $|\nu_i\rangle$ caused by these four reactions with $^{12}\mathrm{C}$ as
\be
r_{i,^{12}\mathrm{C}}(t_d,E_\nu)=\mathrm{N}_{^{12}\mathrm{C}}\sum_{X,\beta} F^{}_{\nu_i}(t^{}_{\rm d}, E^{}_\nu)\cdot |U^{}_{\beta i}|^2 \cdot \sigma_{X}(E_{\nu_\beta})
\label{nuc12rate}
\ee
where the subscript $X$ stands for the four reactions in \refer{c12cc} and \refer{c12nc}.

Further integrating the event spectrum \refer{eratediff} over the neutrino energy, we obtain the total event rate of elastic neutrino-proton scattering
\bea
R^{}_{i, {\rm PS}}(t^{}_{\rm d}) = \int_{E^{\rm min}_\nu}^\infty r_{i,{\rm PS}}(t^{}_{\rm d}, E^{}_\nu) \dd E^{}_\nu \label{erdef} \; ,
\eea
and likewise $R^{}_{i, {\rm ES}}$ for the neutrino-electron scattering \refer{neratediff} and neutrino-$^{12}$C reaction \refer{nuc12rate}. In the case of elastic neutrino-electron scattering, the observed recoil energy should also be larger than $0.2~{\rm MeV}$ for the scintillator detector, but $3.5~{\rm MeV}$ for the water-Cherenkov detector. This implies that the minimal neutrino energy $E^{\rm min}_\nu$ will be different for these two types of detectors. If the target is composed of water, the fiducial mass of 2.5 megaton corresponds to $N^{}_p \approx 2 \times 10^{35}$ and $N^{}_e \approx 1\times 10^{36}$. For the scintillator detector of the same fiducial mass, the proton and electron numbers are quite similar. The number of $^{12}$C, assuming that the liquid scintillator is chosen as linear alkyl bencene ($C_{18}H_{30}$) as in Juno \cite{An:2015jdp}, can be calculated as $N_{^{12}\mathrm{C}}\simeq 3 N_e/23$.

\begin{figure}[!t]
\includegraphics[width=0.52\textwidth]{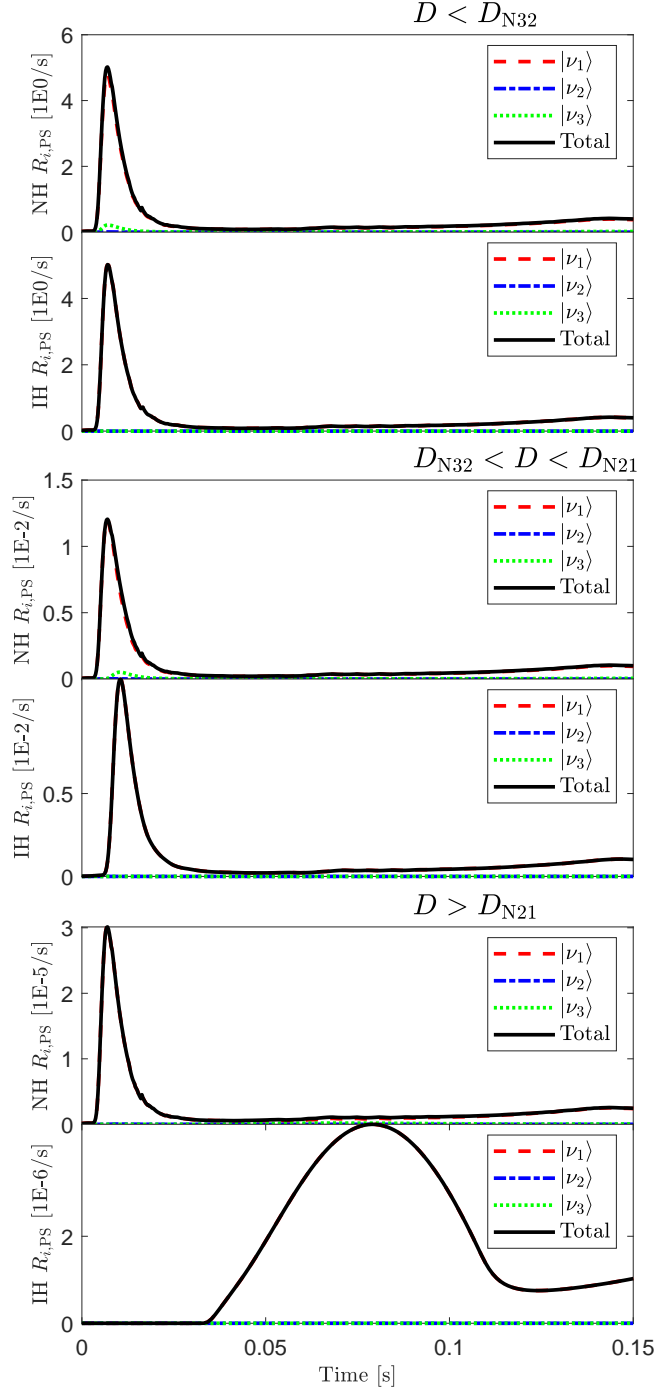}
\caption{\label{figpeaks2mhsim_ps} The neutrino event rates of elastic neutrino-proton scattering for a 2.5 megaton liquid-scintillator detector, where the individual contributions from $|\nu^{}_i\rangle$ (for $i = 1, 2, 3$) are denoted by colored dashed lines and the total rate by black solid line. Three SN distances $D^{}_1 = 1~{\rm Mpc} < D^{}_{\rm N32}$ (top panel), $D^{}_{\rm N32} < D^{}_2 = 20~{\rm Mpc} < D^{}_{\rm N21}$ (middle panel) and $D^{}_3 = 400~{\rm Mpc} > D^{}_{\rm N21}$ (bottom panel) are considered, where the upper and lower plot in each panel corresponds respectively to NH and IH. The lightest neutrino (i.e., $|\nu^{}_1\rangle$ for NH and $|\nu^{}_3\rangle$ for IH) is taken to be massless.}
\end{figure}

\begin{figure}[!t]
\includegraphics[width=0.52\textwidth]{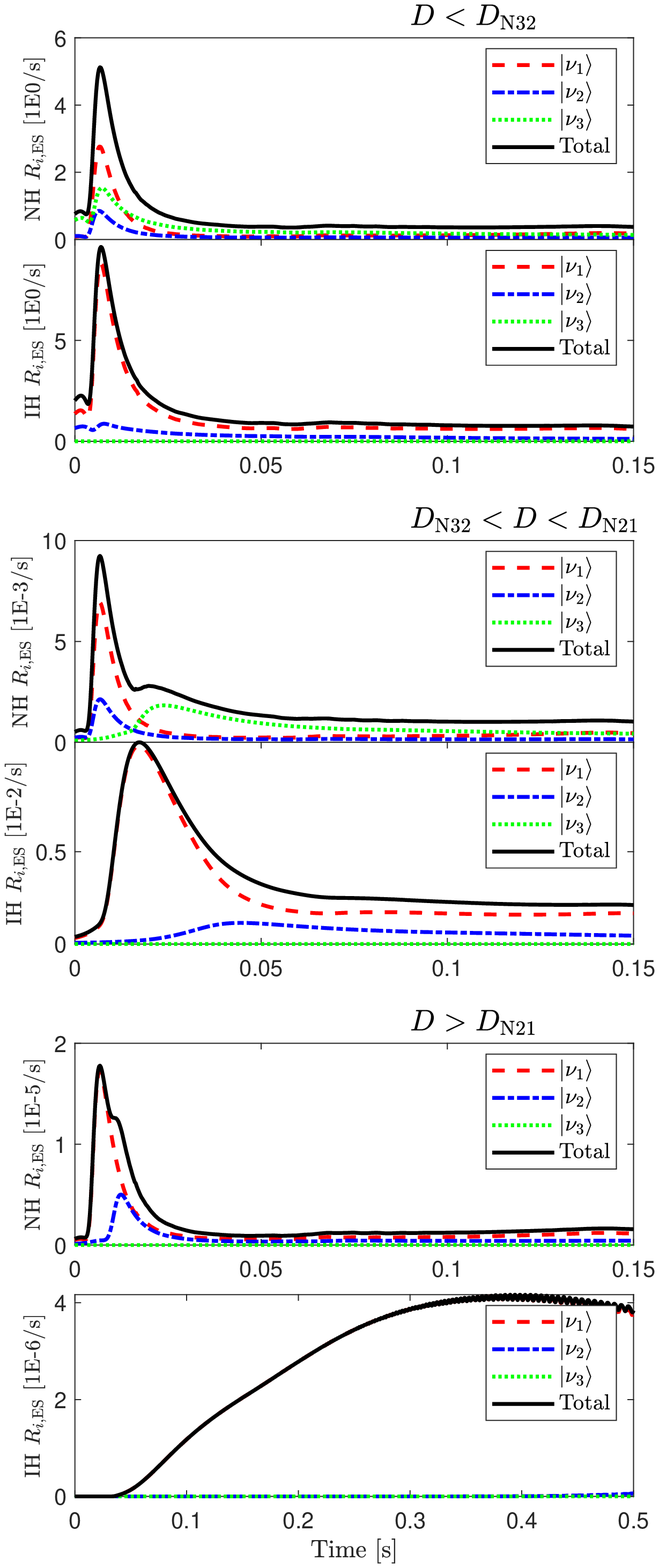}
\caption{\label{figpeaks2mhsim_es} The neutrino event rates of elastic neutrino-electron scattering for a 2.5 megaton liquid-scintillator detector, where the notations and input are the same as those in Fig.~\ref{figpeaks2mhsim_ps}.}
\end{figure}

\begin{figure}[!t]
\includegraphics[width=0.52\textwidth]{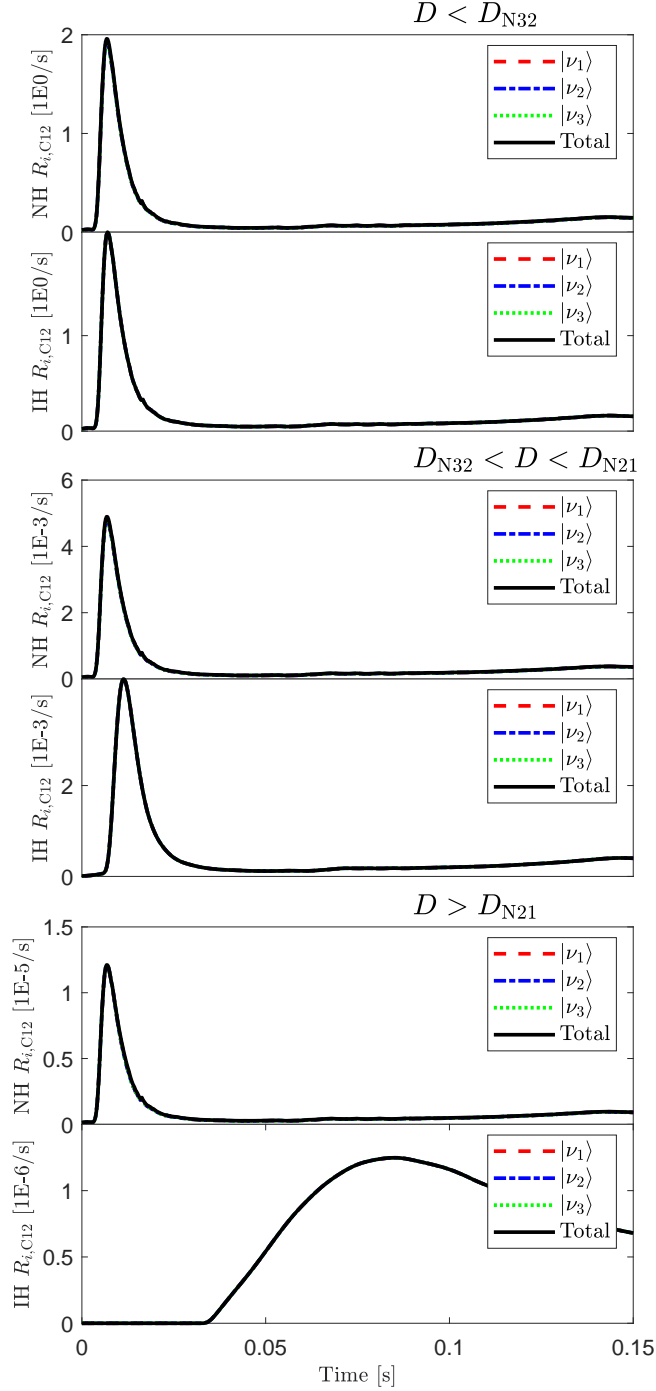}
\caption{\label{figc12rates} The neutrino event rates of neutrino-$^{12}$C reactions for a 2.5 megaton liquid-scintillator detector, where the notations and input are the same as those in Fig.~\ref{figpeaks2mhsim_ps}.}
\end{figure}

In Fig.~\ref{figpeaks2mhsim_ps}, we present the numerical results of the individual event rates $R^{}_{i,{\rm PS}}$ ($i = 1,2,3$) for three neutrino mass eigenstates and also the total rate (the thick and black curve) for either NH or IH. Similarly, the numerical results of $R^{}_{i,{\rm ES}}$ are shown in Fig.~\ref{figpeaks2mhsim_es} and that of $R_{i,^{12}\mathrm{C}}$ in Fig. \ref{figc12rates}. For all these reaction channels, three representative distances, $D^{}_1 = 1$ Mpc for the small distance, $D^{}_2 = 20$ Mpc for the intermediate distance and $D^{}_3 = 400$ Mpc for the long distance, have been considered. In addition, the transition probabilities $P_{ei}(E)$ given in Fig.~2 of Ref.~\cite{Dasgupta:2008cd} have been adopted in the calculations. Some comments on the results are in order.

First of all, let us recapitulate the fractions of neutrino mass eigenstates after the action of both MSW matter effects and collective oscillations on an initial flux of pure $|\nu^{}_e\rangle$ in a SN model from Ref.~\cite{Dasgupta:2008cd}. The fractions depend crucially on the neutrino energy and their main features can be summarized as follows: (1) In the NH case, we have only $|\nu^{}_1\rangle$ for $E \gtrsim 17~{\rm MeV}$, $|\nu^{}_2\rangle$ for $15~{\rm MeV} \lesssim E \lesssim 17~{\rm MeV}$ and $|\nu^{}_3\rangle$ for $E \lesssim 15~{\rm MeV}$; (2) In the IH case, there is a critical energy $E \approx 12~{\rm MeV}$, below which only $|\nu^{}_2\rangle$ survives while only $|\nu^{}_1\rangle$ above. Then, we look at the event rates of neutrino-proton scattering shown in Fig.~\ref{figpeaks2mhsim_ps}. One common feature of all the plots is that the contributions from $|\nu^{}_2\rangle$ and $|\nu^{}_3\rangle$ are negligibly small. This can be well understood by noticing that the observed recoil energy of the final-state proton should be larger than $0.2~{\rm MeV}$, indicating that only the neutrino states with energies above $E > 21.7~{\rm MeV}$ contribute. For both NH and IH, only $|\nu^{}_1\rangle$ meets this requirement, explaining why $|\nu^{}_1\rangle$ dominates the contributions to the total event rate. As an immediate consequence of this observation, there will never appear two or three peaks, which is evident from all the plots in Fig.~\ref{figpeaks2mhsim_ps}. Therefore, it is difficult to tell the difference between NH and IH for the small and intermediate distances. However, for the large distance $D^{}_3 = 400~{\rm Mpc} > D^{}_{\rm N21}$, the $|\nu^{}_1\rangle$ peak in the IH case is broadened significantly compared to that in the NH case. The reason is simply that the lightest neutrino mass has been set to be vanishing in the numerical calculations, namely, the absolute mass $m^{}_1 \approx 49~{\rm meV}$ of $|\nu^{}_1\rangle$ in the IH case is much larger than that $m^{}_1 = 0$ in the NH case.

Now we turn to the event rates of elastic neutrino-electron scattering depicted in Fig.~\ref{figpeaks2mhsim_es}. Since the recoil energy of the final-state electron is not quenched in the liquid scintillator, all the neutrino mass eigenstates can contribute to the event rates. However, the $|\nu_3\rangle$ peaks in the IH case are highly suppressed due to a tiny transition probability $P^{}_{e3}$. Some discussions about this reaction channel are helpful.
\begin{itemize}
\item As neutrinos in the entire energy range contribute, the magnitudes of the event rates of neutrino-electron scattering in all cases of MH and distances are comparable to or sometimes larger than those of the neutrino-proton scattering, in spite of the smaller cross section of neutrino-electron scattering in Eq.~\refer{nuxecs} compared to that in Eq.~\refer{nuallpcs}. In addition, the cross section for electron neutrinos is about six times larger than that of other neutrino flavors, so the neutrino mass eigenstate that has the largest component of $|\nu^{}_e\rangle$ is most important. This non-universality of the cross sections results in different total event rates for the two MHs. Such a difference is most apparently seen by comparing the heights of the total rate peaks in the two subplots in the small distance case of $D^{}_1 = 1~{\rm Mpc} < D^{}_{\rm N32}$. However, the SN distance is too short for three peaks to be well separated in the NH case.

\item For the intermediate distance ($D^{}_{\mbox{\scriptsize N32}}< D^{}_2 = 20~{\rm Mpc} < D^{}_{\mbox{\scriptsize N21}}$), we see that for the NH the event peak due to $|\nu^{}_3\rangle$ is already separated from and lags behind the peaks of $|\nu^{}_1\rangle$ and $|\nu^{}_2\rangle$, while the latter two are still stacked together. For the IH, again the tiny conversion probability $P^{}_{e3}$ completely suppresses the $|\nu^{}_3\rangle$ peaks and therefore there is mainly one combined $|\nu_1\rangle+|\nu_2\rangle$ peak left. The apparent difference of these two subplots  makes it possible to tell the MH for the intermediate distance case when the conversion probabilities inside the supernova follows that in {\it Case (A)} of Sec.~\ref{nbosc}. Indeed, even if the probabilities have to be changed, as long as they are known to a good accuracy such that the heights and shapes of the $|\nu^{}_3\rangle$ peak and $|\nu^{}_1\rangle + |\nu^{}_2\rangle$ peak for the NH and IH can be calculated, the MH can always be deduced by comparing them with the observed ones.

\item For the large distance ($D^{}_3 = 400~{\rm Mpc} > D^{}_{\mbox{\scriptsize N21}}$), one can observe that in the NH case, the peaks due to $|\nu^{}_1\rangle$ and $|\nu^{}_2\rangle$ are already separated while the $|\nu^{}_3\rangle$ peak (not shown in order to make the $|\nu^{}_1\rangle$ and $|\nu^{}_2\rangle$ peaks clear) lags behind them by about 0.4 second. In the IH case, the $|\nu^{}_1\rangle$ peak is also separated from the $|\nu^{}_2\rangle$ peak although the latter is now flattened too much to be seen as a peak. Therefore, the difference between the NH and IH event rates is also quite obvious to determine the MH in the large distance case. Compared to the intermediate distance case, however, the large distance case suffers the drawback of a much lower total event rate.
\end{itemize}

For a 2.5 megaton water-Cherenkov or liquid-scintillator detector and an intermediate distance, for which the MHs can be distinguished, integrating the event rate shown in the middle panel of Fig.~\ref{figpeaks2mhsim_es}, we find that there will be $5.25\times 10^{-4}$ and $1.02\times 10^{-3}$ events in the NH and IH cases, respectively. Even for a one gigaton detector, there will be a total of $0.21$ and $0.41$ events during the first 0.5 second. If we assume the SN distance to be $D^{}_{\mbox{\scriptsize N21}} \approx 133~{\rm Mpc}$, a detector of 450 gigaton is required to register one neutrino event. This is certainly beyond the scope of any current and near future neutrino detectors.

For the neutrino reactions with $^{12}$C that are allowed by liquid-scintillator detectors, the combined event rates of all four reactions in \refer{c12cc}-\refer{c12nc} are shown in Fig. \ref{figc12rates} for three typical distances. It is seen by comparing with the event rates of proton scattering in Fig. \ref{figpeaks2mhsim_ps} that generally, the shape of the event rates of the combined neutrino-$^{12}$C reaction is quite similar to that of the neutrino-proton scattering, although the magnitude of the former is only $\sim 1/3 - \sim 1/2$ of the latter. This is understandable because the cross-section of neutrino-proton scattering is about 1.2$\sim$1.5 times the total cross-section of neutrino-$^{12}$C reactions, and the proton density is 1.5 times the $^{12}$C density. Moreover,  while the neutrino-proton cross-section is universal to all neutrino flavors, the neutrino-$^{12}$C cross-sections (in particular the charge-current interaction \refer{c12cc}) are not. Also similar to the case of proton scattering, the main contribution from the total event rate is also from $|\nu_1\rangle$ for the same reason as in Fig. \ref{figpeaks2mhsim_ps}. Moreover, this also implies that there will not appear two or three peaks in the event rate of the neutrino-$^{12}$C reaction channel and therefore difficult to tell the neutrino mass ordering.

\subsection{Further Discussions \label{nbfactors}}

Finally, we discuss how the variations of a few important parameters affect the event rates. These key parameters include the transition probabilities $P^{}_{ei}(E)$, the initial neutrino spectrum $n^{}_{\nu_e^{}}(t, E)$ and the absolute mass of the lightest neutrino. Here we give some brief remarks on the impact of those parameters and leave a full analysis for future works.

\subsubsection{Transition Probabilities}

\begin{figure}[!t]
\includegraphics[width=0.52\textwidth]{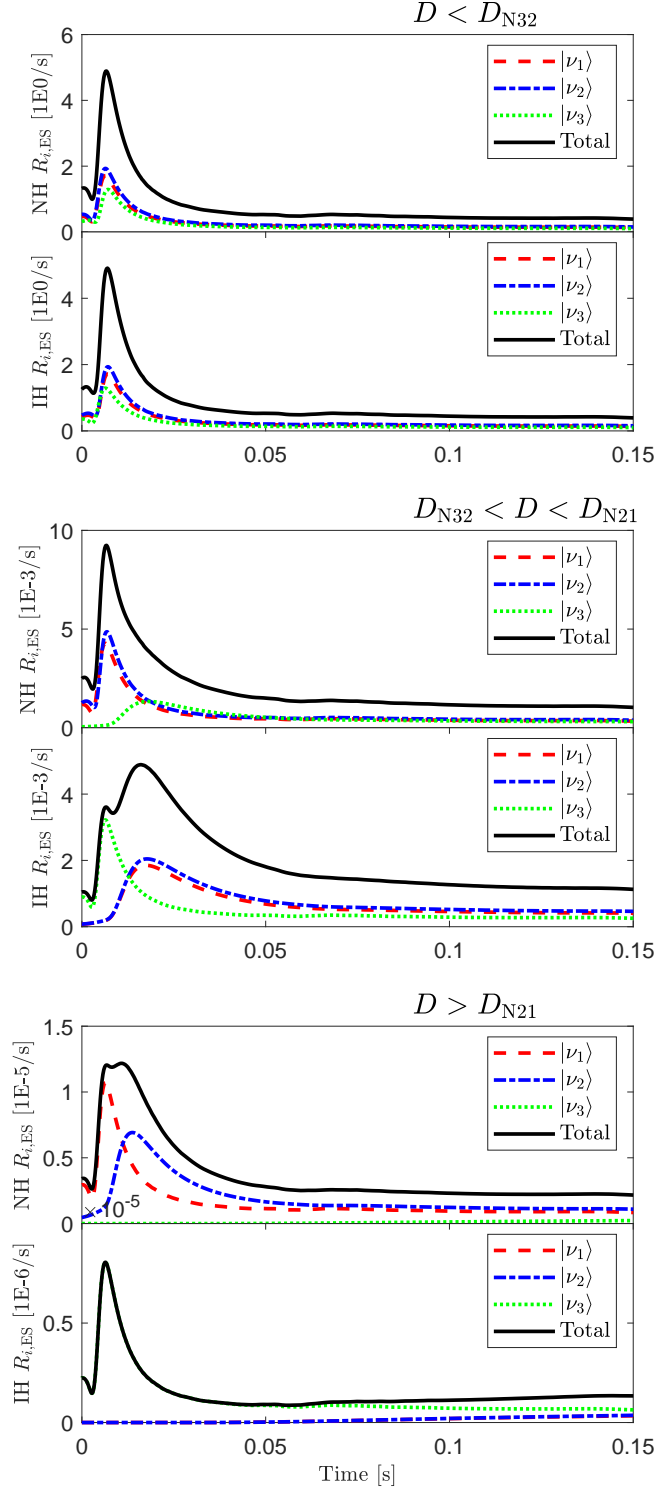}
\caption{\label{figpeaks2mhsimequalprob} The neutrino event rates of elastic neutrino-electron scattering for a 2.5 megaton liquid-scintillator detector, where the notations and input are the same as those in Fig.~\ref{figpeaks2mhsim_es} except that the transition probabilities $P^{}_{ei}$ are replaced by those in {\it Case (B)} of Sec.~\ref{nbosc}.}
\end{figure}

For comparison, we adopt the transition probabilities in {\it Case (B)} of Sec.~\ref{nbosc} and present the event rates in Fig.~\ref{figpeaks2mhsimequalprob}, where the other input parameters are taken to be exactly the same as in Fig.~\ref{figpeaks2mhsim_es}. The main purpose for such a comparison is to show that the feasibility of our method does not depend too critically on these conversion probabilities. For the small distance, the total event rates for both NH and IH are almost identical and thus cannot be used to pin down the MH, as expected. Now the difference is that the probability $P^{}_{e3}$ in the IH case is not small anymore and therefore the $|\nu_3\rangle$ peaks are no longer suppressed. For the intermediate distance, the shapes and arriving order of the $|\nu^{}_1\rangle+|\nu^{}_2\rangle$ peak and the $|\nu^{}_3\rangle$ peak are clearly seen different for the two MHs, and therefore can also be used to tell the MH. For the large distance case, since this time in the IH case the $|\nu^{}_3\rangle$ is not suppressed by the probability, the $|\nu^{}_3\rangle$ peak is behind and ahead of the $|\nu_1\rangle$ peak by about 0.4 second for the NH and IH, respectively.

\subsubsection{Initial Spectrum}

The initial spectrum Eq.~\refer{nbn} is expected to describe the true neutrino spectrum quite well~\cite{Tamborra:2014hga}. Substituting it into Eq.~\refer{neratediff} we can obtain the event spectrum and then study the impact of two pivotal parameters, namely, $\langle E(t)\rangle$ and $E^{}_{\mbox{\scriptsize rms}}(t)$. If the average energy is increased, while other quantities are kept unchanged, the central energy of spectrum will shift to a larger value while its peak value will decrease due to the constraint from the fixed total luminosity. However, considering the energy dependence of the cross section, we find that a higher average energy will in general enhance the total event rate $R^{}_i$.

The parameter $E^{}_{\mbox{\scriptsize rms}}(t)$ controls how much the spectrum is concentrated on the central energy. Its influence on both the neutrino spectrum and the event rate in Eq.~\refer{neratediff} are secondary compared to $\langle E(t)\rangle$. It turns out that as $E^{}_{\mbox{\scriptsize rms}}(t)$ decreases, the total event rate $R^{}_i$ also drops down but slowly. Besides its impact on the total event rate, a better concentration of the neutrinos in energy also means a narrower spread in the arrival time after including the time delay effect. Given that the main difficulty of our method here comes from the resolution of different peaks, a narrower distribution of the initial neutrino spectrum in energy should be more favorable.

The luminosity $L(t)$ affects the resolvability of the peaks of neutrino mass eigenstate through its appearance in the spectrum in Eq.~\refer{nbn}. As we have seen earlier, if the neutronization peak duration is prolonged by a factor of $N$, then we need the SN distance to be increased roughly by the same factor in order to achieve the same temporal separation of the peaks. Consequently, the total event rate will be reduced by a factor of $N^2$. As demonstrated in Ref.~\cite{Lentz:2011aa, Kuroda:2012nc}, a more complete inclusion of relevant weak interactions and the full general relativistic treatment tend to increase the width of the neutronization burst. On the other hand, Ref.~\cite{Buras:2005tb, Serpico:2011ir, Liebendoerfer:2003es} suggest that the SN for a smaller progenitor mass (e.g., around 10 to 15 solar masses) will do the opposite, i.e., producing a neutronization neutrino burst narrower in time. The latter observation indicates that the SNe with smaller progenitor masses are favored for our method to work effectively.

\subsubsection{Absolute Neutrino Masses}

In the numerical results presented in Figs.~\ref{figpeaks2mhsim_ps}, ~\ref{figpeaks2mhsim_es} and ~\ref{figpeaks2mhsimequalprob}, we have assumed the lightest neutrino to be massless. Therefore, the shape of the event spectrum of the lightest neutrino resembles exactly the feature of neutronization burst from the numerical simulation data~\cite{Tamborra:2014hga}. A nonzero mass of the lightest neutrino mass eigenstate leads to a time delay, and accordingly the luminosity peak will be widened in time and its maximum will be lowered, since the total neutrino number is kept unchanged. On the other hand, it is now known from the observational data of Baryon Acoustic Oscillation and Planck~\cite{Giusarma:2016phn} that the sum of three neutrino masses has an upper limit, namely, $m^{}_1 + m^{}_2 + m^{}_3 < 0.176~{\rm eV}$, which after taking into account the measured neutrino mass-squared differences leads to an upper bound of $m^{}_1 < 0.052~{\rm eV}$ for NH and $m^{}_3 < 0.044~{\rm eV}$ for IH.

We have checked that even if the upper bounds on the lightest neutrino are saturated and the largest distance of $D^{}_3 = 400~{\rm Mpc}$ is assumed, the width and height of the luminosity peak are changed only by about a factor of 1.5. There exists even more stringent constraints on the sum of three neutrino~\cite{DiValentino:2015sam}, for which we find that the luminosity peak is practically unchanged in both its width and strength compared to those in Figs.~\ref{figpeaks2mhsim_ps}, \ref{figpeaks2mhsim_es} and \ref{figpeaks2mhsimequalprob}. Moreover, the essential idea is to probe the relative shift in time among different mass eigenstates, which are little affected by the total peak shape and strength. Therefore, the impact of absolute neutrino masses on our method can be safely ignored.

\section{Black Hole Formation\label{secbh}}

The method using neutronization neutrino burst to determine the MH is not useful when the SN distance $D$ is not large enough to resolve three peaks of neutrino mass eigenstates. One can observe from Eq.~\refer{dmndef} that $D$ is mainly limited by the relatively large duration $\Delta t^{}_{\rm N} \approx 5~{\rm ms}$ of the neutronization burst. Fortunately, in the scenario of failed SNe, there exists another characteristic process with an even shorter time span: the termination of neutrino signals due to the BH formation during the accretion phase.

The rate of BH formation core-collapse SNe and the formation mechanism are still uncertain, and numerical simulations crucially depend on the initial progenitor mass and the details of models. In many concrete SN models with BH formation, the neutrino signal will be abruptly terminated when the neutrino flux is still measurably high. After the energy-dependent ToF is taken into account, the sharp cutoff on the neutrino flux will cause characteristic signals at the detector with different descending rates, corresponding to different neutrino mass eigenstates. More importantly, the BH formation is a phase transition process that takes very short time $\Delta t^{}_{\rm B}$, which can be estimated as $\Delta t^{}_{\rm B} \lesssim 2R/c \simeq 0.1$ ms, where $R$ is the radius of the active region of neutrino emission. In comparison with $\Delta t^{}_{\rm N}$ for the neutronization burst, $\Delta t^{}_{\rm B}$ is much shorter and therefore allows for more practical SN distances and higher statistics.

For the resolution for the cutoff edges in the signals of three mass eigenstates, a similar criterion to Eq. \refer{dmndef} yields the requirement on the SN distance
\be
D \geq \frac{\Delta t^{}_{\rm B}}{ 5.15~\mbox{ms}}\frac{\left(\langle E \rangle /\mbox{10~MeV}\right)^2}{|\Delta m_{ij}^2|/\mbox{eV}^2}\cdot 10~\mbox{kpc}\equiv D^{}_{\mbox{\scriptsize Bij}} \; . \label{dbhlimit}
\ee
Corresponding to $\Delta m^2_{21}$ and $|\Delta m^2_{32}|$, there will also be two characteristic distances $D^{}_{\mbox{\scriptsize B21}}$ and $D^{}_{\mbox{\scriptsize B32}}$, which take the typical values $D^{}_{\rm B21} \approx 2.58~{\rm Mpc}$ and $D^{}_{\rm B32} \approx 79.3~{\rm kpc}$
for an average neutrino energy $\langle E \rangle = 10~{\rm MeV}$ and $\Delta t^{}_{\rm B} \simeq 0.1$ ms. Apparently, because $\Delta t^{}_{\rm B}$ is about 50 times smaller than $\Delta t^{}_{\rm N}$, the minimal distances for given $\Delta m_{ij}^2$ and $\langle E\rangle$ to resolve the neutrino tail due to the BH formation should be about $2\%$ of that for the neutronization burst. As a consequence, when the distance is the same, there will be a much larger event rate and statistical significance for the MH discrimination.

For clarity, we define the time derivative of the event rate as $\Gamma^{}_i \equiv -\dd R^{}_i/\dd t$, for which the time $t^{\rm max}_i$ to reach its maximum for the neutrino mass eigenstate $|\nu^{}_i\rangle$ will be different from one another. For the BH forming SN at a distance larger than $D^{}_{\mbox{\scriptsize B21}}$, we can find that $t^{\rm max}_i$ for the three mass eigenstates will be temporally separated by a duration larger than $\Delta t^{}_{\rm B}$. For NH, $t^{\rm max}_1$ and $t^{\rm max}_2$ of $|\nu_1\rangle$ and $|\nu_2\rangle$ appear temporally close to each other, but earlier than that of $|\nu_3\rangle$. For IH, the opposite is true, namely, $t^{\rm max}_3$ appears first and is further separated from $t^{\rm max}_1$ and $t^{\rm max}_2$. For the distance between $D^{}_{\mbox{\scriptsize B21}}$ and $D^{}_{\mbox{\scriptsize B32}}$, the locations of $t^{\rm max}_1$ and $t^{\rm max}_2$ will not be resolvable but are still separated from $t^{\rm max}_3$. If the relative strength of the signals for three mass eigenstates is known, then the MH can still be deduced from the shapes of the descending event rates. Similar to the case of neutronization neutrino burst, for $ D < D^{}_{\mbox{\scriptsize B32}}$, again the MH will not be deducible in this way. Taking these distances for example, we perform numerical computations of the event rates in the scenario of BH formation.

\subsection{Neutrino Event Rates}

Just before the BH formation, all the neutrinos and antineutrinos of three flavors can be produced. For the detection of antineutrinos, it is obvious that we should first consider the inverse beta decay (IBD) of $\overline{\nu}^{}_e$, whose cross section for SN neutrinos is much larger than those of other reactions in water or liquid scintillator. For neutrinos, the main contributions are from elastic scattering on protons and electrons. Therefore, we will only take account of these three processes in the our calculations. The IBD event rate of the $i$-th mass eigenstates of antineutrinos $|\overline{\nu}^{}_i\rangle$ is given by
\bea
R^{}_{i,{\rm IBD}} = N^{}_p \int^\infty_{1.8~{\rm MeV}} F^{}_{\overline{\nu}^{}_i} P^{}_{ei} \cdot \sigma^{}_{\rm IBD}(E^{}_\nu) \cdot \dd E^{}_\nu \; , \label{bhner}\eea
while the total rate can be calculated by summing over all possible contributions, namely, $R^{}_i = R^{}_{i, {\rm IBD}} + R^{}_{i, {\rm PS}} + R^{}_{i, {\rm ES}} + R^{}_{i, ^{12}\mathrm{C}}$, where the part of elastic proton scattering should be omitted for a water-Cherenkov detector. A simple approximation to the IBD cross section was given in Ref. \cite{Strumia:2003zx}. However here we will use a more accurate differential cross-section given by Ref. \cite{Vogel:1999zy} which takes account into the weak magnetism effect and nucleon recoil effect
\be
\frac{\dd  \sigma^{}_{\rm IBD}}{\dd \cos\theta}(E^{}_\nu) = \frac{\sigma_0}{2} \left[ (f^2+3g^2)+(f^2-g^2)v_e\cos\theta\right] E_e^{(1)}p_e^{(1)}-\frac{\sigma_0}{2}\left[ \frac{\Gamma}{M}\right] E_e^{(0)}p_e^{(0)} \label{dibd}\ee
where $\theta$ is the angle between the antineutrino and positron direction in the lab frame, $\sigma_0,~f,~g,~f_2$ are some constants, $E_e^{(i)}$ and $p_e^{(i)}$ are the electron energy and momentum to order $i$. $\Gamma$ is the term standing for the correction of the $1/M$ order to the lower order cross section. For simplicity, we refer the readers to Ref. \cite{Vogel:1999zy} for their exact definitions and only point out that $E_e^{(i)},~p_e^{(i)}$ and $\Gamma$ are functions of angle $\theta$ and energy $E_\nu$ of the incoming neutrino. Finally, numerically integrating \refer{dibd} over $\theta$ allows us to find the total IBD cross-section $\sigma^{}_{\rm IBD}(E^{}_\nu)$, which we will direct use in this work. It is worth mentioning that $R^{}_{i,{\rm PS}}$ and $R^{}_{i, {\rm ES}}$ in this section include the contributions from both neutrinos and antineutrinos, for which the cross sections of neutrino-proton scattering in Eq.~\refer{eratediff} are equal while those for neutrino-electron scattering in Eq.~\refer{nuxecs} are different for electron and non-electron flavors. Furthermore, the cross section in  Eq.~\refer{nuxecs} can be adapted for antineutrinos by exchanging $\epsilon^{}_- \leftrightarrow \epsilon^{}_+$~\cite{Marciano:2003eq}.

The fluxes $F^{}_{\overline{\nu}^{}_i}$ and $F^{}_{\nu^{}_i}$ at the time $t^{}_{\rm d}$ on the Earth can be computed from the spectra $n^{}_{\nu_e}(t, E)$, $n^{}_{\overline{\nu}^{}_e}(t, E)$ and $n^{}_{\nu^{}_x}(t, E)$ where $\nu^{}_x$ collectively denote $\nu^{}_\mu$ and $\nu^{}_\tau$ and their anitparticles, multiplying them by the conversion probabilities $P^{}_{\beta i} = |U^{}_{\beta i}|^2$ and taking into account the time delay due to Eq.~\refer{tof}. Explicitly, the formula for $|\overline{\nu}^{}_i\rangle$ can be expressed as
\be
F^{}_{\overline{\nu}_i}(t^{}_{\rm d}, E)= \frac{1}{4\pi D^2} \sum_{\beta} n^{}_{\overline{\nu}_\beta^{}}(t^{}_{\rm d} - \Delta t^{}_{i1},E) P^{}_{\beta i} \; , \label{fcalc2}
\ee
which is very similar to Eq.~\refer{fcalc} except that the contributions from all flavors are summed up and the conversion probabilities $P^{}_{\beta i}$ (for $\beta = e, \mu, \tau$) are energy independent. Sucn an independency is expected because the BH formation normally takes place in the accretion phase, when the self-induced collective oscillations are found to be suppressed by the large matter density~\cite{Chakraborty:2011nf, Chakraborty:2011gd, Sarikas:2011am}. Hence the neutrino spectra experience only adiabatic conversions inside the SN, and then the mass eigenstates freely stream from the SN surface to the Earth.

\begin{figure}[!t]
\includegraphics[width=0.6\textwidth]{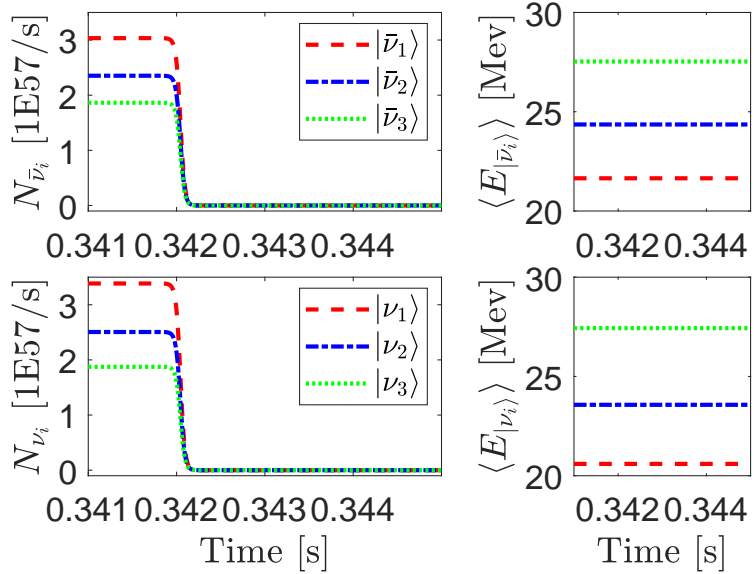}
\caption{\label{figf6flavor} The energy integrated spectra and average energies of three mass eigenstates of antineutrinos (first row) and neutrinos (second row) are given in the left and right panels, respectively, where only a narrow time window is shown for the abrupt termination of (anti)neutrino emission.}
\end{figure}

In our numerical calculations, we use the $n^{}_{\overline{\nu}^{}_e}$, $n^{}_{\nu^{}_e}$ and $n^{}_{\nu^{}_x}$ spectra obtained from the SN simulation in Ref.~\cite{Nakazato:2012qf} for the progenitor star of 30 solar masses. After converting them into the spectra of neutrino mass eigenstates, we find the energy integrated spectra for each $\nu^{}_i$
\be N_{\nu_i }(t)=\int \dd E \sum_\beta n_{\nu_\beta}(t,E)P_{\beta i} \ee
and similarly for each $\overline{\nu}^{}_i$ the corresponding $N_{\overline{\nu}_i}(t)$, which together with their corresponding average energies are given respectively in the left and right panels of Fig.~\ref{figf6flavor}. Since $n^{}_{\nu^{}_e}$ is slightly larger than $n^{}_{\overline{\nu}^{}_e}$, we have the integrated spectra $N^{}_{\nu^{}_i}$ also larger than $N^{}_{\overline{\nu}^{}_i}$ before BH formation. Moreover, the time structure of neutrino signals is mainly determined by the neutrino emission in a very short period right before the BH formation, we can verify that the number spectra are constant in this short time slot before BH forms \cite{Beacom:2000qy}. Some observations from Fig.~\ref{figf6flavor} are summarized below:
\begin{enumerate}
\item The integrated spectra before BH formation decreases in the order of $|\overline{\nu}^{}_1\rangle$, $|\overline{\nu}^{}_2\rangle$ and $|\overline{\nu}^{}_3\rangle$, and that of $|\overline{\nu}^{}_1\rangle$ is roughly 1.5 times that of $|\overline{\nu}^{}_3\rangle$. In addition, the integrated spectra for neutrinos $|\nu^{}_i\rangle$ are almost in the same order as those of the corresponding antineutrino $|\overline{\nu}^{}_i\rangle$, except that the $|\nu^{}_1\rangle$ spectrum is slightly larger than that of $|\overline{\nu}^{}_1\rangle$ due to the higher initial luminosity of $|\nu^{}_e\rangle$ than $|\overline{\nu}^{}_e\rangle$.

\item Regarding the average energies of mass eigenstates shown in the right panels of Fig. \ref{figf6flavor}, note that indeed they are {\it time dependent} as directly extracted from simulation result \cite{Nakazato:2012qf}. They only look like constant because in this small time window responsible for the neutrino tail, their numerical value barely change. It is also notable that although the average energies slightly increase in the order of $|\nu_1\rangle$ or $|\bar{\nu}_1\rangle$ to $|\nu_3\rangle$ or $|\bar{\nu}_3\rangle$, the average energies of mass eigenstates are all lying between 21 MeV to 28 MeV, approximately with an average value of 24 MeV. Using Eq.~\refer{dbhlimit}, one can estimate the two critical distances
  $D_{\mbox{\scriptsize B21}}\simeq 14.9~\mbox{Mpc and }D_{\mbox{\scriptsize B32}}\simeq 457~\mbox{kpc}$. These changes are obviously due to the enhancement of the average energies.
\end{enumerate}
\begin{figure}[!t]
\includegraphics[width=0.7\textwidth]{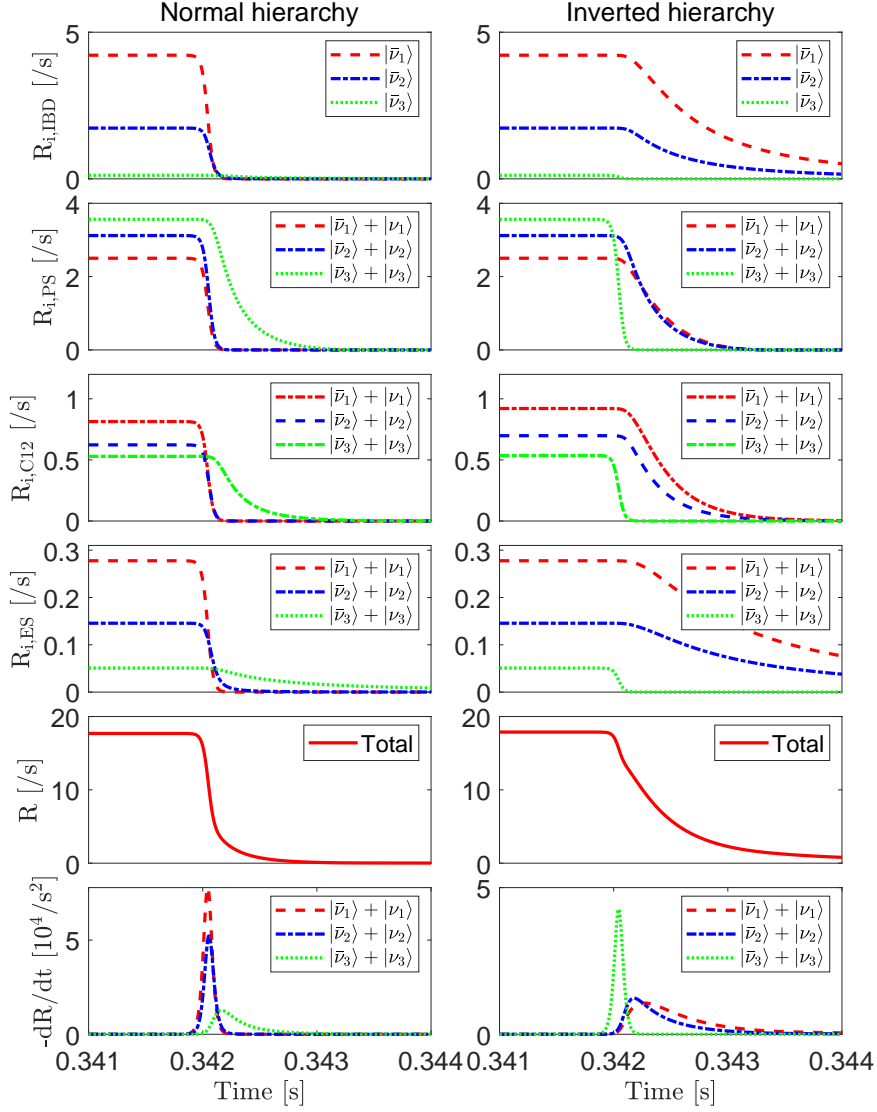}
\caption{\label{figbh2mhsim} Numerical results of neutrino events for $D = 5~{\rm Mpc}$ and a 2.5 megaton detector in the scenario of the BH formation, where the plots from the first to third row correspond to IBD, neutrino-proton elastic scattering and neutrino-electron elastic scattering, respectively. The total event rate $R(t)$ and its time derivative $-\dd R(t)/\dd t$ are given in the last two rows. The plots in the left column are for NH while those in the right for IH. }
\end{figure}

In Fig.~\ref{figbh2mhsim} we show the event rates around the time of the termination of the neutrino signal due to BH formation, where the SN distance is $D = 5$ Mpc and a 2.5 megaton liquid-scintillator detector is assumed. The results for NH and IH are presented in the left and right column, respectively. Some comments on the numerical results are in order.
\begin{itemize}
\item From the first row of Fig.~\ref{figbh2mhsim}, we present the IBD events for both hierarchies, where one can see that the $|\nu^{}_1\rangle$ rate roughly doubles that of $|\nu^{}_2\rangle$ which is much higher than the $|\nu^{}_3\rangle$ rate. From Fig. \ref{figf6flavor}, it is already clear that the fluxes of $|\overline{\nu}^{}_1\rangle$ and $|\overline{\nu}^{}_2\rangle$ are the same and both are twice that of the $\overline{\nu}^{}_3$ flux, so the IBD event rates are largely determined by the conversion probabilities from $|\overline{\nu}^{}_i\rangle \to |\overline{\nu}^{}_e\rangle$. For the neutrino-proton scattering event rates shown in the second row, because of the universality of the cross section for all six mass eigenstates, only the energy dependence of the cross section in Eq.~\refer{eratediff} and the average energies will cause differences among the event rates. As indicated in Fig.~\ref{figf6flavor}, the average energies of $|\nu^{}_3\rangle$ and $|\overline{\nu}^{}_3\rangle$ are the highest, leading to their neutrino-proton scattering event rates comparable to those of the $|\nu^{}_1\rangle$ and $|\overline{\nu}^{}_1\rangle$, and $|\nu^{}_2\rangle$ and $|\overline{\nu}_2\rangle$. The third and fourth rows show the detection rate of $^{12}$C reaction for liquid scintillator detector and of neutrino-electron scattering respectively. Unlike the neutronization burst case in Figs.~\ref{figpeaks2mhsim_ps} and \ref{figpeaks2mhsim_es} where the events rates of electron and proton scatterings are comparable, here the former is smaller by more than one order of magnitude than the latter. The reason is simply that the average energies of neutrinos range from 21 MeV to 28 MeV, and most neutrinos have energies beyond the cut-off energy 22 MeV. In contrast, the average neutrino energy of the neutronization burst is only about 12 MeV.

\item The total event rates, and the its time derivatives for each mass eigenstate are given respectively in the last two rows of Fig.~\ref{figbh2mhsim} for both NH and IH. We can see that the total rates at the time long before and after the BH formation are equal for both hierarchies. It should be noticed that the signal decays for heavy mass eigenstates (i.e., $|\nu^{}_1\rangle$ and $|\nu^{}_2\rangle$ and their antiparticle states for IH while $|\nu^{}_3\rangle$ and $|\overline{\nu}^{}_3\rangle$ for NH) are much slower than those for light mass eigenstates. This is the key feature and can be used to distinguish one neutrino MH from another if enough neutrino events can be detected. Such a difference between NH and IH can also be clearly seen from the time derivative of event rates shown in the last row. For the NH, the peak of $|\nu^{}_3\rangle +|\overline{\nu}^{}_3\rangle$ is low in height and slowly decreasing, and comes later than the higher peaks of    the first and second mass eigenstates. For the IH, the opposite happens: the $|\nu_3\rangle+|\bar{\nu}_3\rangle$ peak becomes sharp and decreases fast while located on the right of the peaks of other mass eigenstates.
\end{itemize}
In the above numerical computations, the SN distance $D$ is set to 5 Mpc, which is larger than $D^{}_{\mbox{\scriptsize B32}} \simeq 1$ Mpc, the distance required for the maximal descending rates of $|\nu^{}_3\rangle$ and $|\nu^{}_2\rangle$ to separate by a gap larger than $\Delta t^{}_{\rm B}$, obtained by using $\langle E \rangle = 24$ MeV, as shown in Fig.~\ref{figf6flavor}. With the total event rate in Fig.~\ref{figbh2mhsim}, one can easily calculate that for $D = 5~{\rm Mpc}$ and a detector of 2.5 megaton there will be just about $(3 \sim 6)\cdot 10^{-3}$ events during the entire period of 0.2 ms. At the distance $D_{\mbox{\scriptsize B32}} \simeq 1~{\rm Mpc}$ and for one gigaton detector, we can obtain roughly 60 events that may be statistically adequate to distinguish the MHs.

\subsection{Further Discussions}

Now we discuss how the input parameters affect the tail of neutrino signals from the BH forming SN. Compared to the case of neutronization burst, the situation for the BH case is much simpler. First, the variation of neutrino spectra with time is practically constant because the time window under discussion is narrow enough. Second, the flavor conversions of neutrinos and antineutrinos are adiabatic, and thus it is straightforward to figure out the transition probabilities. Third, the impact of the absolute mass of the lightest mass eigenstate, as argued in the case of neutronization burst, is expected to be small if the cosmological bound on the sum of neutrino masses is applied. Therefore, we analyze the effect of the distance $D$ on the total event rate in the following.
\begin{figure}[!t]
\includegraphics[width=0.7\textwidth]{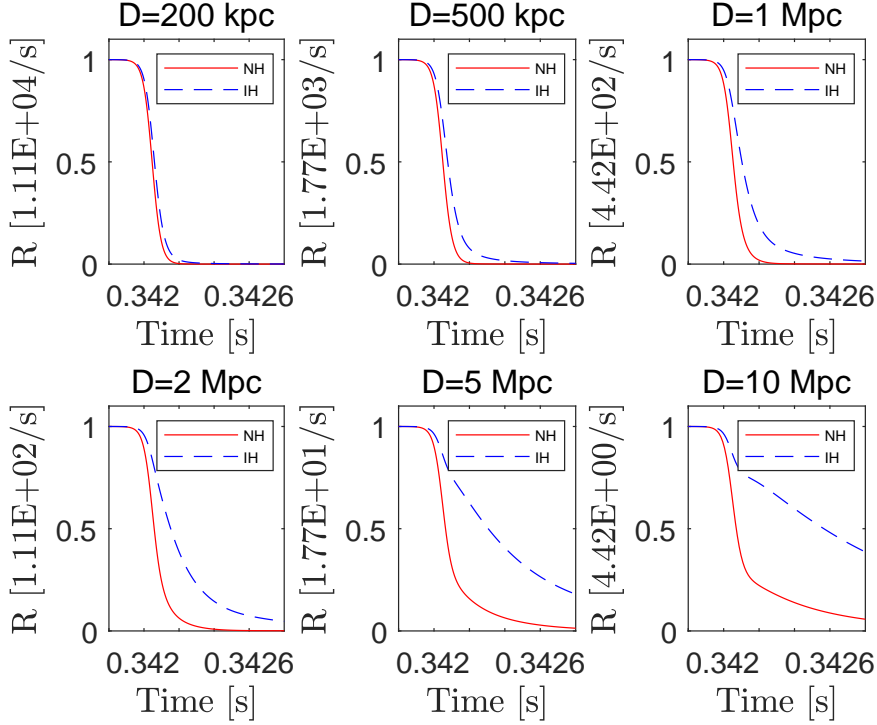}
\caption{\label{figbhmanyd} The total event rates of neutrinos from the BH forming SNe at different distances, where a detector of 2.5 megaton is assumed. }
\end{figure}

The first and most important effect of the SN distance is that the total event rate during the decay of neutrino signals is inversely proportional to $D^2$. Moreover, the distance also has an influence on the shape of the total event rate, which turns out to be crucial for the MH discrimination. In Fig.~\ref{figbhmanyd}, we plot the total event rates as functions of time for a series of distances ranging from 200 kpc to 10 Mpc for the two MHs. In fact, it is more practical to determine the distance by using other methods, e.g., optical observations, and then the MH can be determined by comparing the measurement with the predicted signal shape. For the previous example, once the distance is determined (e.g., $D = 1$ Mpc), then the neutrino MH can be deduced by a comparison of the descending rates $\dd R/\dd t$ of both hierarchies.

\section{Concluding Remarks\label{secdis}}

In this work, we have carried out a phenomenological analysis of the ToF effects of massive neutrinos, and applied them to the SN neutrinos with short-time characteristics. For the neutronization $\nu^{}_e$ burst of core-collapse SNe, the peaks corresponding to different neutrino mass eigenstates for the NH will appear in a temporal order different from that for the IH. A clear discrimination between two MHs requires a large distance for the SN, whereas a high statistics favors SNe at small distances. For this reason, we have found that it seems impossible to determine the neutrino MH via the neutronization burst for a typical core-collapse SN in the currently running and near future SN neutrino detectors. However, for the BH forming SNe, the abrupt termination of neutrino emission is shown to allow for a determination of neutrino MH with an enough statistics. In this case, a liquid-scintillator detector of one gigaton will register about 60 events for a SN located at 1 Mpc during the period of BH formation.

The impact of neutrino flavor conversions (particularly those in the case of neutronization burst), absolute neutrino masses and the supernova distance (particularly in the case of BH forming SNe) on the discriminating power is also discussed. It is shown that neutrino flavor conversions generally will not destroy the feasibility of the basic idea as long as the flavor conversions do not completely suppress the peaks in such a way that only one mass eigenstate is left. As for the absolute mass of the lightest neutrino, it has been numerically verified that if the cosmological upper bound on the sum of neutrino masses is applied, the shapes of the neutrino event rates in both neutrinonization burst and BH formation cases will have little changes. The SN distance does affect the shape of the neutrino signal decay in the case of the BH forming SN. In reality, one can fix the SN distance by other means and then extract the information of neutrino MH from observations.

There are a few other factors that are related to the applicability of our conclusions. The most important one is the uncertainty in the luminosities and spectra of SN neutrinos. For the neutronization burst, the uncertainties involved in the numerical simulations of SN explosions can introduce noticeable differences in the duration time of the burst and its luminosities. Moreover, the simulations do not take into account the full neutrino-neutrino interactions which severely affect the emerging neutrino spectra. If this burst is narrower, with a higher luminosity and more $|\overline{\nu}_3\rangle$ component, then the approach of using the neutronization burst to probe the MH becomes more promising. Secondly, the fraction of the BH forming SNe among all core-collapse SNe can vary from a few percent to a sizable value (see, e.g., Ref.~\cite{Beacom:2000qy}), and the neutrino luminosities and spectra before the BH formation could also be quite different~\cite{Fischer:2008rh}. The value that we have used for the luminosity of each flavor is on the order of $10^{52}~{\rm erg}~{\rm s}^{-1}$, which comes from a SN with BH formation at its early stage \cite{Beacom:2000qy}. If the BH is formed at a later stage, then the luminosity and thus the total event rate can be one order of magnitude lower. One more factor associated with SN neutrino detection is about the low supernova probability within applicable distance. It was estimated that the core collapse SN rate will be only $3.2^{+7.3}_{-2.6}$ in the Galaxy per century \cite{Adams:2013ana} and that within 1 Mpc is only about 0.04$\sim$0.08 per year \cite{Ando:2005ka}. These numbers suggest that other experimental methods might potentially discriminate the neutrino mass hierarchies earlier than any new supernova neutrino observation.

Besides the neutronization burst and the BH termination of neutrinos, we emphasize that the basic idea of using the ToF difference to distinguish the MH is also applicable to any neutrino sources with short-time characteristics. Here we considered only two such characteristics, i.e., the neutronization neutrino $|\nu_e\rangle$ burst and neutrino spectrum tail during black hole formation, because these are the only two {\it theoretically} known features with short time duration in SN neutrino spectrum. Given that there was effectively only one SN neutrino observation (SN1987A) and the large uncertainty in simulations, the true detailed SN neutrino spectrum with high statistics is still experimentally unknown. Therefore there remains possibility that some other features with similar or even shorter time duration might exist in the spectrum. If so, these features will also be usable to determine the neutrino MH in the same way. However, until then, to more accurately validate the idea in this work for SN neutrinos, we will have to perform more careful studies of the initial neutrino fluxes from realistic SN simulations including a full treatment of neutrino flavor conversions, which will be left for future works.

\begin{acknowledgments}
The authors are grateful to Jia-Shu Lu and Yu-Feng Li for helpful discussions. This work was supported in part by the National Natural Science Foundation of China under Grant Nos. 11504276 and 11547310 (J.J. and Y.W.), by the National Recruitment Program for Young Professionals and the CAS Center for Excellence in Particle Physics (S.Z.).
\end{acknowledgments}

\end{document}